\documentclass[twocolumn,superscriptaddress,prb,amsmath,amssymb,showpacs,lengthcheck,reprint]{revtex4-1} 
\usepackage[T1]{fontenc}
\usepackage[latin9]{inputenc}
\usepackage{mathrsfs}
\usepackage{graphicx}
\usepackage{color}

    \newcommand{\note}[1]{\textcolor{red}{#1}}
    
\usepackage{braket}

\begin{document}

\title{Performance of quantum-dot-based tunnel-injection lasers: A theoretical analysis} 
\author{M. Lorke}
\affiliation{Institute for Theoretical Physics, University of Bremen, 28359 Bremen, Germany}

\author{S. Michael}
\affiliation{Institute for Theoretical Physics, University of Bremen, 28359 Bremen, Germany}

\author{M. Cepok}
\affiliation{Institute for Theoretical Physics, University of Bremen, 28359 Bremen, Germany}

\author{F. Jahnke}
\affiliation{Institute for Theoretical Physics, University of Bremen, 28359 Bremen, Germany}

\begin{abstract}
Tunnel-injection lasers promise advantages in modulation bandwidth and temperature stability in comparison to conventional laser designs. 
In this paper, we present results of a microscopic theory for laser properties of tunnel-injection devices and a comparison
to a conventional quantum-dot laser structure.  
In general, the modulation bandwidth of semiconductor lasers is affected by the steady-state occupations of electrons and holes via the presence of spectral hole burning.
For tunnel-injection lasers with InGaAs quantum dot emitting at the telecom wavelength of 1,55$\mu$m, we demonstrate
that the absence of spectral hole burning favors this concept over conventional quantum-dot based lasers.
\end{abstract}

\maketitle

\section{Introduction}

Semiconductor laser devices are important components for fiber-optical communication. Requirements for opto-electronic applications include low threshold current, 
high temperature stability, and large modulation bandwidth. 
In conventional quantum dot (QD) laser devices, the pump process generates carriers in delocalized states, 
while the QD ground state is used for the carrier recombination into the laser mode.

\begin{figure}[!ht]
\centering
\includegraphics[trim=2cm 4.5cm 1cm 1.4cm,clip,scale=0.39,angle=0]{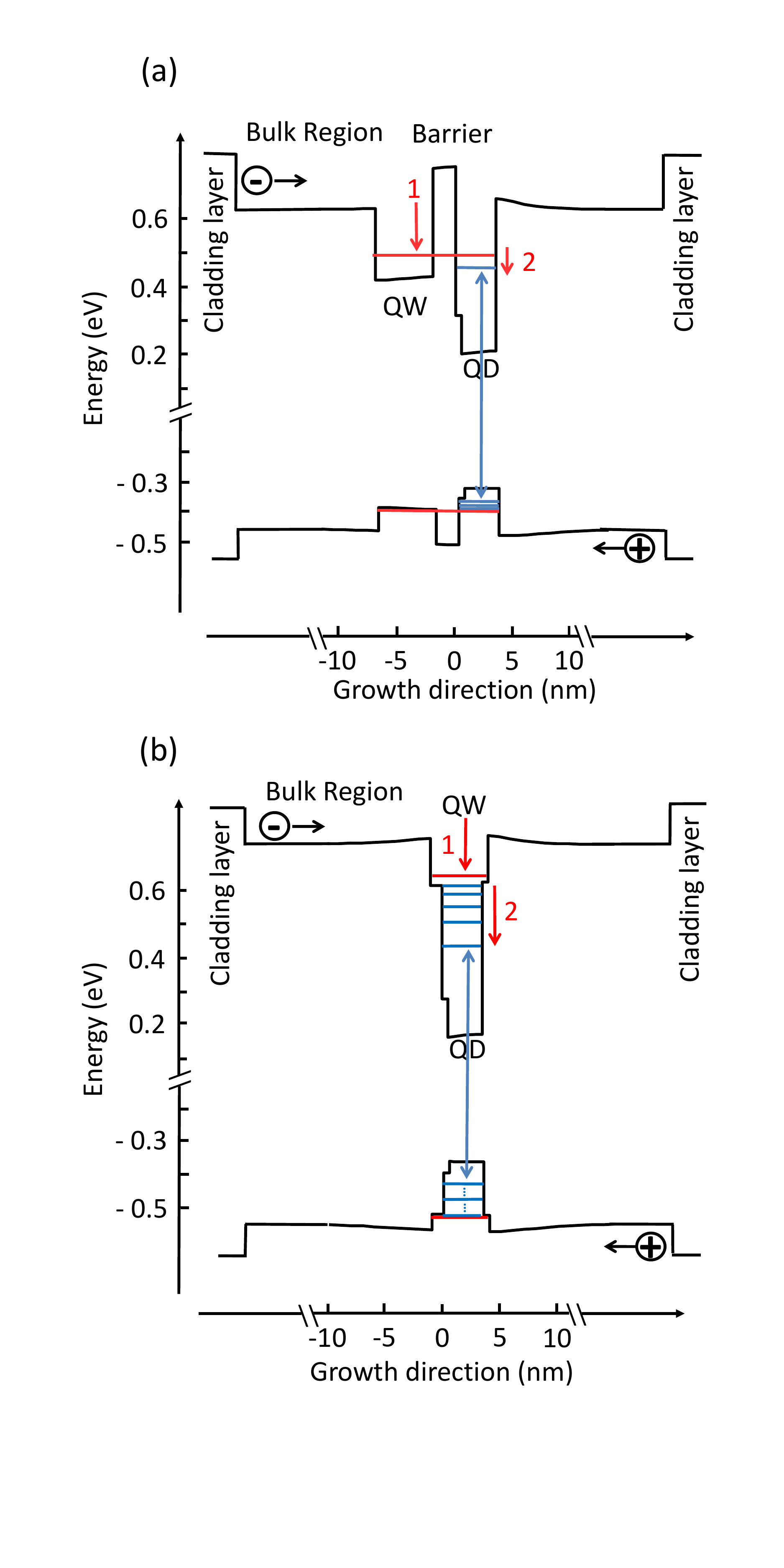}
\caption{(a) Band structure of a TI QD device with the IW separated by a thin barrier from the QDs that provide the laser level.
Process 1 and 2 (red arrows) describe the capture of conduction band electrons from the bulk states into the IW and the relaxation 
from the hybridized IW-QD state into the optically active QD ground state, respectively.
(b) DWELL structure used for comparison. The QD emits at the same wavelength. 
Here, the carrier dynamics is determined by the capture from the bulk into the QW, 
followed by capture from the QW into the QDs and intra-QD relaxation.
\label{fig1}}
\end{figure}

Tunnel injection devices have been proposed to enhance both temperature stability and modulation properties \cite{bhattacharya2002tunnel,fathpour2003linewidth,bhattacharya2003carrier,bhowmick2014high}. 
The latter are limited by hot carrier effects, a problem that the tunnel injection scheme is designed to overcome by feeding cold carriers from an injector well (IW) 
directly to the optically active QD states \cite{bhattacharya2002tunnel,fathpour2003linewidth}. This concept has been demonstrated for QDs \cite{bhattacharya2002tunnel,pavelescu2009high,bhowmick2014high} as well as for quantum well systems \cite{bhattacharya1996tunneling,zhang19970}. 
In recent experiments with devices utilizing the TI scheme 
improvements of GaAs-QD based high-power lasers \cite{pavelescu2009high}
and ultra-fast gain recovery \cite{pulka2012ultrafast} have been achieved.
Despite these successes, open questions remain regarding the design requirements. 
To which extent needs the LO-phonon resonance to be considered in the level alignment? 
What is the role of non-equilibrium carrier effects?
What are suitable designs to suppress hot carrier effects?
To address these points we provide a theoretical analysis that connects
the electronic states and carrier scattering processes on one hand and the 
resulting laser properties such as temperature stability and modulation bandwidth on the other hand.

\begin{figure*}[!ht]
\centering
\includegraphics[trim=3.5cm 4cm 3cm 2cm,clip,scale=0.52,angle=0]{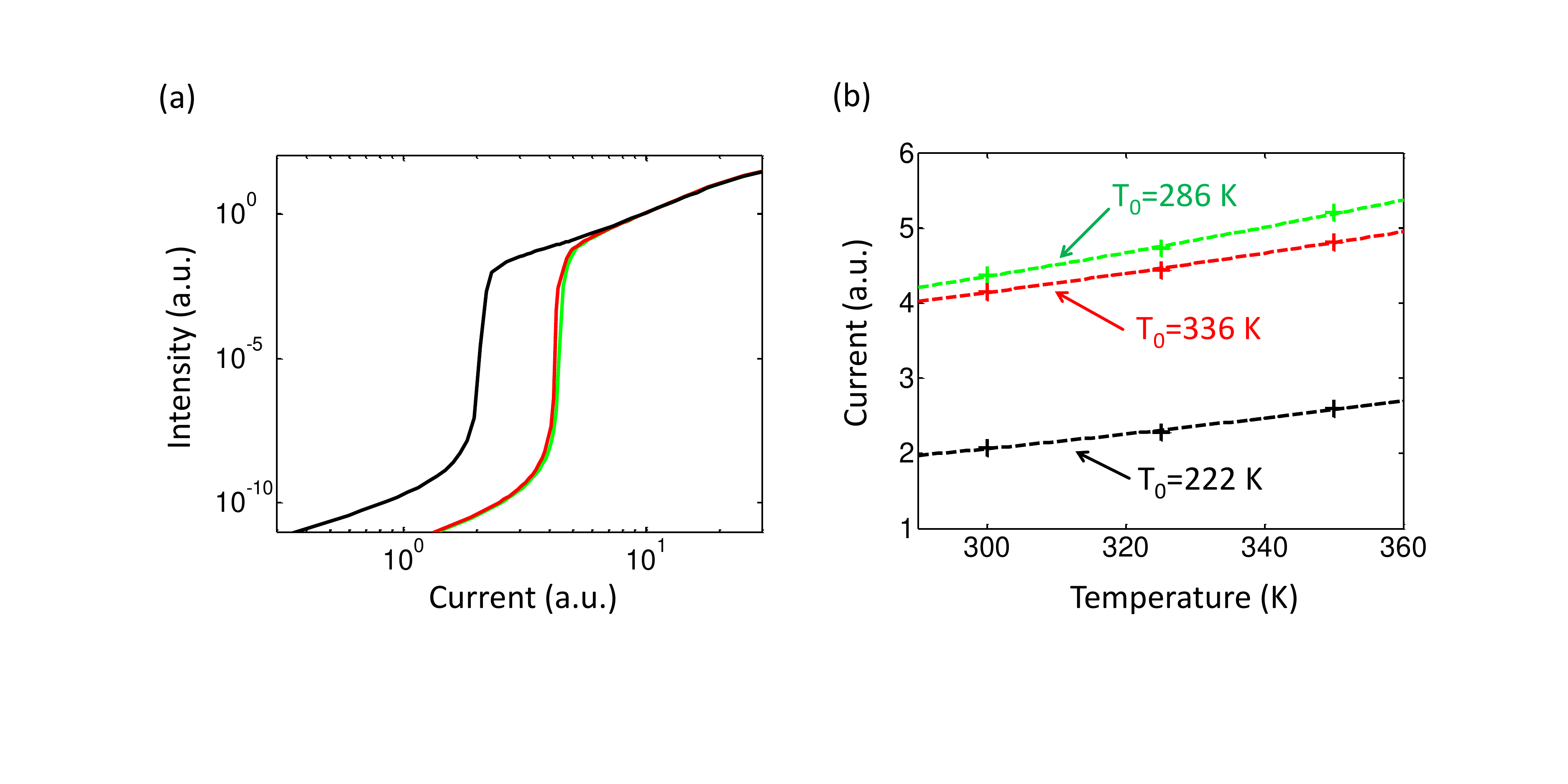}
\caption{Steady state properties of the DWELL and two different TI lasers. (a) Input-Output characteristics for the DWELL structure (black) and TI 
devices with QD sizes of 16x24nm (green) and 16x32nm (red), respectively. Dimensions refer to the in-plane extensions of ellipsoidal QDs.
(b) Temperature dependence of these three devices for a temperature range of 300K-350K. The parameter $T_0$ stems from an exponential fit of the temperature dependent 
threshold current (see text). 
\label{fig2}}
\end{figure*}

In an earlier publication \cite{Michael:18}, we discussed the physics of carrier scattering in tunnel injection structures. 
When the design leads to hybridized states between IW and QD, this significantly enhances the carrier capture into the laser levels.
In this work, we analyze the laser properties of TI devices on a microscopic footing and discuss the 
advantages of the TI design.
On a general level it is known \cite{Norris,Blood} that the temperature dependence of the laser threshold is influenced by the temperature dependence of the carrier scattering.
We show for the TI system that the carrier scattering is rather insensitive to temperature in the investigated parameter range, which in turn improves the temperature stability
of the laser emission.

Additionally an enhanced hole occupation of the IW improves temperature stability in a similar manner to p-doping of conventional QD laser structures \cite{Smowton}.
We find that a main limiting factor of the small signal modulation bandwidth is spectral hole burning, that leads to a strongly nonlinear gain.
In TI devices this spectral hole burning is largely suppressed, as the carrier dynamics is fast 
enough to ensure sufficient carrier supply into the optically active states
due to the high density of states in the IW and the efficient hybridization of IW and QD states.
The latter provides enhanced carrier scattering to sustain a quasi-equilibrium situation, even under laser operation conditions.

\section{Results}

To assess the advantages of TI lasers, we perform a comparison to a conventional QD-based laser design in the form of a dot-in-a-well (DWELL) structure.
As shown in previous work \cite{Michael:18}, the design of the QDs can affect the hybridization strength and hence the efficiency of carrier scattering.
Therefore two different QD geometries for the TI devices are considered, one with moderate (16x24nm) and one with near-optimal (16x32nm) hybridization strength.
These geometries are within the size distribution recently found in HRTEM investigations of TI-QD laser devices \cite{Schowalter:priv_comm}. For the comparison, we ensure that 
both TI and DWELL structures operate at the same emission wavelength (1550nm) and closely resemble devices currently under experimental investigation.


\begin{figure*}[!ht]
\centering
\includegraphics[trim=6cm 9cm 6cm 6cm,clip,scale=0.42,angle=0]{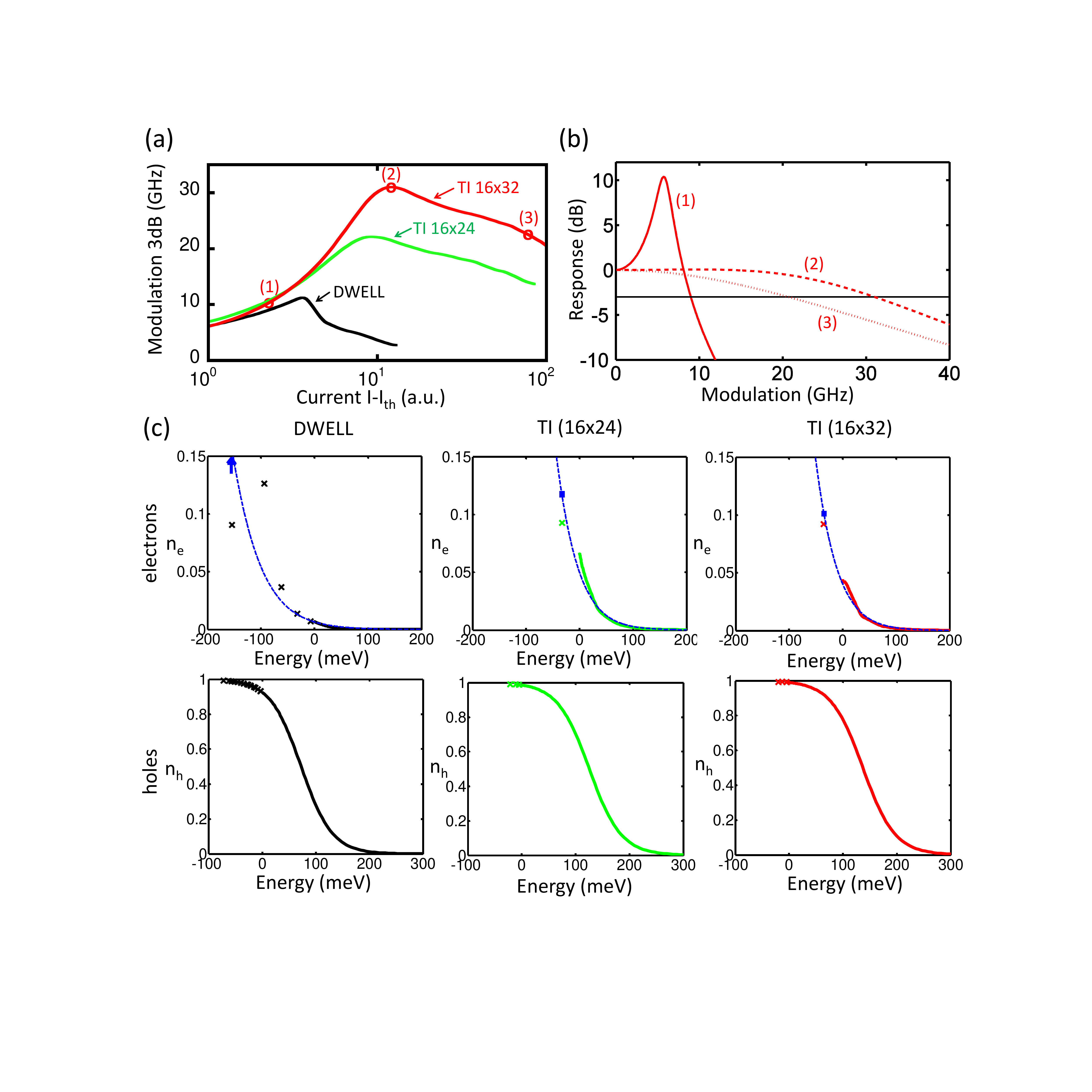}
\caption{(a) 3dB modulation frequency as function of current for all three devices. The color coding corresponds to Fig.~\ref{fig2}. (b) Modulation response spectra for the 16x32nm QD device and three different currents
as marked in (a). (c) Electron (top row) and hole (bottom row) populations for the current with the maximum 3dB frequency for the DWELL structure (left column), 
the 16x24nm TI structure (middle column) and the 16x32nm TI structure (right column). 
The solid lines denote the occupation function of the continuum of states, while the crosses show the QD occupations. Thermal distributions for the electrons are depicted as guide to the eye
by the blue lines.
\label{fig3}}
\end{figure*}

\begin{figure}[!ht]
\centering
\includegraphics[trim=1cm 2cm 5cm 3cm,clip,scale=0.32,angle=0]{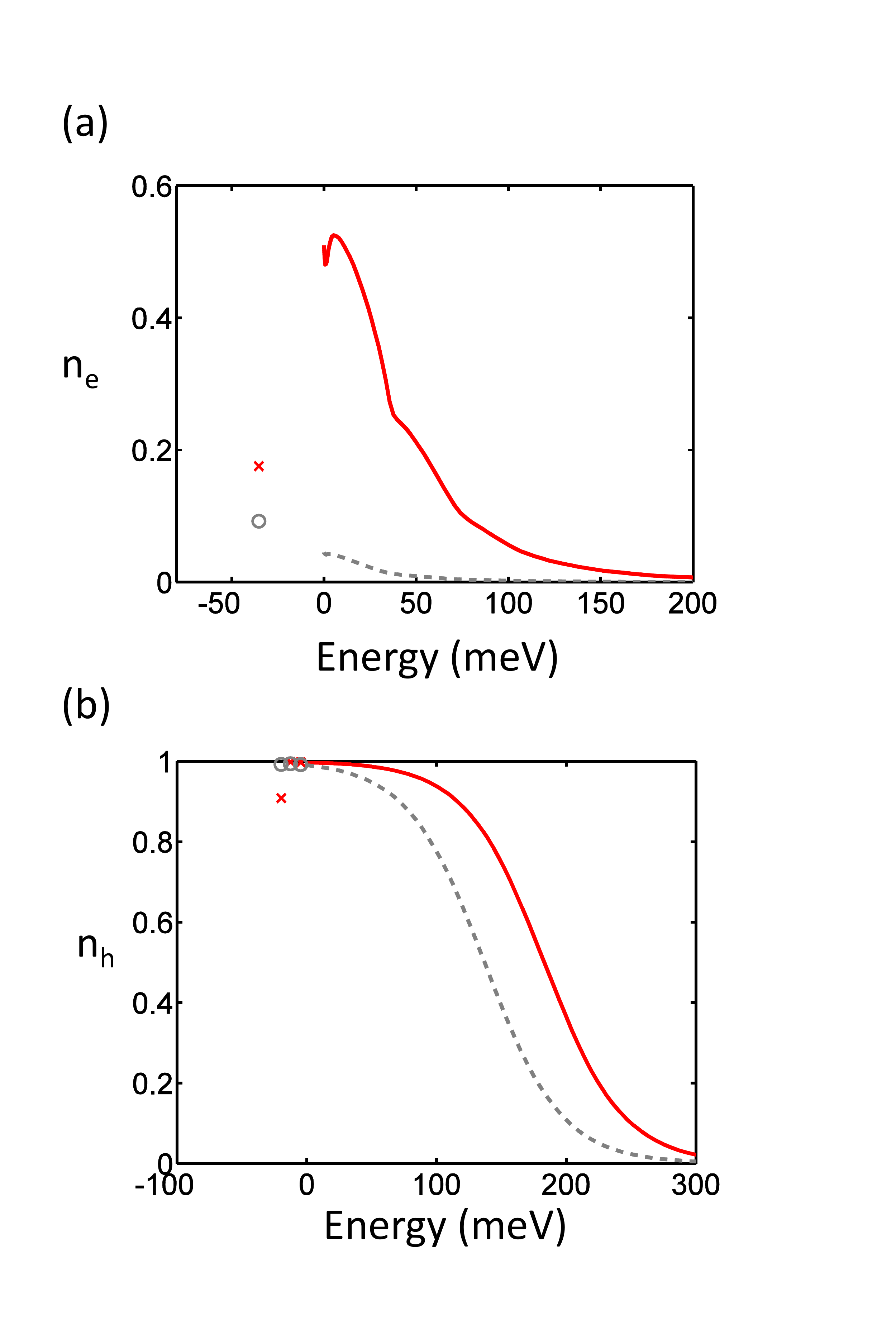}
\caption{Electron (left) and hole (right) populations as function of energy for the injection current with maximal modulation bandwidth (grey dashed line) 
and at a higher injection current (red solid line). The crosses denote the QD populations. The injection currents correspond to labels (2) and (3) in Fig.~\ref{fig3}(a). The results are obtained for the 16x32nm TI structure.
\label{fig4}}
\end{figure}

As described in the supporting information, the corresponding band-structures are shown 
in Fig.~1, obtained from $\vec{k}\cdot\vec{p}$ theory using the nextnano 3 package \cite{nextnano3}. 
The band-bending near the QD and IW results from the included strain field. The small edge of the potential next to the QD represents to the wetting layer, which, however,
carries no bound states due to its small thickness. 
For details of our theoretical model we refer to the supporting information.
In Fig.~2(a) input-output characteristics for DWELL and TI structures are shown. 
To focus our comparison on the changes inherent to the TI design, we use the same non-radiative decay rate in both systems. 
Under this assumption, we find a factor of 2 lower threshold current for the DWELL device. 
For our comparison, the excited carrier population of the TI well is somewhat larger than that of the DWELL 
continuum states as a result of the energetic distance to the QD laser levels.
Assuming the same non-radiative rate, this introduces stronger non radiative losses for the TI structure. We emphasize that this outcome is specific for this type of comparison.
In general, the actual non-radiative losses in particular devices have a stronger influence than the differences between TI and DWELL.

In Fig.~2(b) the temperature dependence of the threshold current is analyzed. 
The $T_0$ values stem from a fit with the exponential dependence $I_\text{th}(T)=I_0\exp\frac{T}{T_0}$
and are in good agreement with those found experimentally for TI devices \cite{bhattacharya2003carrier,bhowmick2014high}. 
Both TI devices clearly show superior temperature stability compared to the DWELL device. 
The carrier scattering via the hybridized state \cite{Michael:18} in the TI structure
is much less temperature sensitive than in the DWELL structure, where the relaxation into the lasing state occurs via a cascade (process 2 in Fig 1 (b)). 
We also find that the temperature stability of the TI device with near-optimal hybridization conditions (16x32nm) surpasses that of the TI laser with moderate hybridization (16x24nm)
Additionally, similar to the discussion in Ref.~\cite{Smowton}, the temperature changes of the gain in high-population scenarios are reduced.

As a figure of merit for the modulation performance of the laser devices, the 3dB-bandwidth \cite{Coldren:95} is provided in Fig.~\ref{fig3}(a).
Both TI structures clearly outperform the DWELL structure by a factor of 2 for moderate hybridization efficiency (16x24nm) and by a factor of 3 for near-optimal hybridization conditions (16 x 32nm). 
For all devices, the modulation bandwidth shows a distinct maximum about an order of magnitude above threshold current \cite{Abdollahinia:18} and is 
reduced significantly for higher injection currents. 
To understand this decrease, individual modulation spectra are shown in Fig.~\ref{fig3}(b). 
For high currents the damping of relaxation oscillations is increased by an increase in carrier scattering rate, leading to a decrease in the 3dB bandwidth. 
This behavior is analogous to the situation found in QD-based nanolasers \cite{Lorke:11}.

To analyze the origin of the higher modulation speed of TI structures, we show in Fig.~\ref{fig3}(c) the electron and hole occupations
for the maximum 3dB frequency of the respective device. 
In the DWELL structure (left panel) a distinct non-thermal behavior for the electrons due to spectral hole burning is observed. As a result, for
the maximum of modulation speed the carrier supply into the lasing state is insufficient. 
In contrast, for the TI structures the electron populations are much closer to a quasi-equilibrium situation even in the case with moderate hybridization efficiency (16x24nm). 
This is caused by the much higher density of states in the IW compared to the DWELL structure, where the optically active state is primarily filled by higher lying QD states.
As the hybridization is more efficient and the carrier scattering is faster for the 16x32nm structure, 
even less deviation from a thermal situation is found and hence the maximum 3dB bandwidth is increased further.

This leads to the following picture. 
In TI structures, the presence of the \note{besser erklaeren} cold IW carrier reservoir, together with fast carrier scattering 
between hybridized IW-QD states \cite{Michael:18} allows for laser operation without significant spectral hole burning.
As this reduces the gain nonlinearity, a significant improvement of the small signal modulation properties is obtained. 
This improvement is influenced by the hybridization efficiency and is therefore sensitive to the QD morphology and IW material composition (cf. Fig. 5 in Ref.~\cite{Michael:18}), but not to a 
tuning to the LO-phonon energy as suggested previously \cite{bhattacharya2003carrier}. The optimum of the modulation speed (for a particular device) is reached
if the carrier scattering is fast enough to sustain a quasi-equilibrium situation.

In Fig.~\ref{fig4} we analyze the decrease of modulation bandwidth with increasing current further by showing the population functions for QD and IW states at the 
maximum of the modulation speed and at an elevated current as depicted in Fig.~\ref{fig3}(a) by labels (2) and (3), respectively.
At the maximum of modulation the carrier distribution for the electrons in the IW is nearly thermal 
which is indicative of sufficient carrier supply.

For an increased current the populations become non-thermal, as the QD population has a much lower value than the energetically higher lying continuum.
This is indicative of pronounced spectral hole burning.
As for the DWELL structure (cf. Fig.~\ref{fig3}(c)) this non-thermal population is connected with an increase of gain nonlinearity and a loss of modulation bandwidth.
The dip in electron occupation near the IW band edge marks the energy for which the strongest hybridization and hence 
the most efficient carrier scattering occurs. As before, the non-thermal character leads to a significant degradation of the 3dB bandwidth.

This result also illustrates a deviation from quantum well or bulk laser devices. In the QD-based lasers investigated here, the \emph{total} carrier density 
does not clamp with injection currents above the threshold current. While the clamping is observed \emph{approximately} for the optically active state,
the density in the delocalized continuum states continues to rise with increasing injection current. 
Hence also detrimental effects like non-radiative decay, that depend on the \emph{total} carrier density rather than on the carrier
density within the QD ground state increase above threshold.

\section{Conclusion}
In this work, we have investigated the physical mechanism behind the increased temperature stability and modulation bandwidth of TI-based lasers.
We find that the increase in temperature stability is governed by carrier scattering being less temperature dependent in the TI structures compared to the DWELL device.
The modulation bandwidth is enhanced in comparison to conventional QD laser devices as the TI scheme is able to provide sufficient 
carrier supply to sustain a quasi-equilibrium situation without significant spectral hole burning. Our results show that the bandwidth 
can be increased by designing the QD and IW in a way that hybridization is efficient.
We do not find any evidence of a necessity to tune the energy levels to the LO-phonon energy.

\begin{acknowledgments}
The authors would like to thank J.~P.~Reithmaier (U. Kassel) and G.~Eisenstein (Technion) for fruitful discussions. 
We acknowledge funding from the DFG and a grant for CPU time from the HLRN (Hannover/Berlin).
\end{acknowledgments}




\begin{thebibliography}{17}
\providecommand{\natexlab}[1]{#1}
\providecommand{\url}[1]{\texttt{#1}}
\expandafter\ifx\csname urlstyle\endcsname\relax
  \providecommand{\doi}[1]{doi: #1}\else
  \providecommand{\doi}{doi: \begingroup \urlstyle{rm}\Url}\fi

\bibitem[Bhattacharya and Ghosh(2002)]{bhattacharya2002tunnel}
P~Bhattacharya and S~Ghosh.
\newblock Tunnel injection in 0.4 ga 0.6 as/gaas quantum dot lasers with 15 ghz
  modulation bandwidth at room temperature.
\newblock \emph{Applied physics letters}, 80\penalty0 (19):\penalty0
  3482--3484, 2002.

\bibitem[Fathpour et~al.(2003)Fathpour, Bhattacharya, Pradhan, and
  Ghosh]{fathpour2003linewidth}
S~Fathpour, P~Bhattacharya, S~Pradhan, and S~Ghosh.
\newblock Linewidth enhancement factor and near-field pattern in tunnel
  injection in/sub 0.4/ga/sub 0.6/as self-assembled quantum dot lasers.
\newblock \emph{Electronics Letters}, 39\penalty0 (20):\penalty0 1443--1445,
  2003.

\bibitem[Bhattacharya et~al.(2003)Bhattacharya, Ghosh, Pradhan, Singh, Wu,
  Urayama, Kim, and Norris]{bhattacharya2003carrier}
P.~Bhattacharya, S.~Ghosh, S.~Pradhan, J.~Singh, Z.-K. Wu, J.~Urayama, K.~Kim,
  and T.~B. Norris.
\newblock Carrier dynamics and high-speed modulation properties of tunnel
  injection ingaas-gaas quantum-dot lasers.
\newblock \emph{IEEE Journal of Quantum Electronics}, 39\penalty0 (8):\penalty0
  952--962, 2003.

\bibitem[Bhowmick et~al.(2014)Bhowmick, Baten, Frost, Ooi, and
  Bhattacharya]{bhowmick2014high}
Sishir Bhowmick, Md~Zunaid Baten, Thomas Frost, Boon~S Ooi, and Pallab
  Bhattacharya.
\newblock High performance inas/in$_{0.53}$ga$_{0.23}$al$_{0.24}$as/inp quantum
  dot 1.55$\mu$ tunnel injection laser.
\newblock \emph{IEEE Journal of Quantum Electronics}, 50\penalty0 (1):\penalty0
  7--14, 2014.

\bibitem[Pavelescu et~al.(2009)Pavelescu, Gilfert, Reithmaier, Martin-Minguez,
  and Esquivias]{pavelescu2009high}
Emil-Mihai Pavelescu, C~Gilfert, Johann~P Reithmaier, A~Martin-Minguez, and
  Ignacio Esquivias.
\newblock High-power tunnel-injection 1060-nm ingaas--(al) gaas quantum-dot
  lasers.
\newblock \emph{IEEE Photonics Technology Letters}, 21\penalty0 (14):\penalty0
  999--1001, 2009.

\bibitem[Bhattacharya et~al.(1996)Bhattacharya, Singh, Yoon, Zhang,
  Gutierrez-Aitken, and Lam]{bhattacharya1996tunneling}
Pallab Bhattacharya, J~Singh, H~Yoon, Xiangkun Zhang, A~Gutierrez-Aitken, and
  Yeeloy Lam.
\newblock Tunneling injection lasers: A new class of lasers with reduced hot
  carrier effects.
\newblock \emph{IEEE journal of quantum electronics}, 32\penalty0 (9):\penalty0
  1620--1629, 1996.

\bibitem[Zhang et~al.(1997)Zhang, Gutierrez-Aitken, Klotzkin, Bhattacharya,
  Caneau, and Bhat]{zhang19970}
X.~Zhang, A.~Gutierrez-Aitken, D.~Klotzkin, P.~Bhattacharya, C.~Caneau, and
  R.~Bhat.
\newblock 0.98-/spl mu/m multiple-quantum-well tunneling injection laser with
  98-ghz intrinsic modulation bandwidth.
\newblock \emph{IEEE Journal of Selected Topics in Quantum Electronics},
  3\penalty0 (2):\penalty0 309--314, 1997.

\bibitem[Pulka et~al.(2012)Pulka, Piwonski, Huyet, Houlihan, Semenova, Lematre,
  Merghem, Martinez, and Ramdane]{pulka2012ultrafast}
Jaroslaw Pulka, Tomasz Piwonski, Guillaume Huyet, John Houlihan, E~Semenova,
  A~Lematre, Kamel Merghem, Anthony Martinez, and Abderrahim Ramdane.
\newblock Ultrafast response of tunnel injected quantum dot based semiconductor
  optical amplifiers in the 1300 nm range.
\newblock \emph{Applied Physics Letters}, 100\penalty0 (7):\penalty0 071107,
  2012.

\bibitem[Michael et~al.(2018)Michael, Lorke, Cepok, Carmesin, and
  Jahnke]{Michael:18}
Stephan Michael, Michael Lorke, Marian Cepok, Christian Carmesin, and Frank
  Jahnke.
\newblock Interplay of structural design and interaction processes in
  tunnel-injection semiconductor lasers, 2018.

\bibitem[Norris et~al.(2005)Norris, Kim, Urayama, Wu, Singh, and
  Bhattacharya]{Norris}
{T. B.} Norris, K.~Kim, J.~Urayama, {Z. K.} Wu, J.~Singh, and {P. K.}
  Bhattacharya.
\newblock Density and temperature dependence of carrier dynamics in
  self-organized ingaas quantum dots.
\newblock \emph{Journal Physics D: Applied Physics}, 38\penalty0 (13):\penalty0
  2077--2087, 7 2005.
\newblock ISSN 0022-3727.
\newblock \doi{10.1088/0022-3727/38/13/003}.

\bibitem[Blood et~al.(1988)Blood, Colak, and Kucharska]{Blood}
P.~Blood, S.~Colak, and A.~I. Kucharska.
\newblock Temperature dependence of threshold current in gaas/algaas quantum
  well lasers.
\newblock \emph{Applied Physics Letters}, 52\penalty0 (8):\penalty0 599--601,
  1988.
\newblock \doi{10.1063/1.99647}.

\bibitem[Smowton et~al.(2007)Smowton, Sandall, Mowbray, Liu, and
  Hopkinson]{Smowton}
P.~M. Smowton, I.~C. Sandall, D.~J. Mowbray, Hui Liu, and Matthew~N Hopkinson.
\newblock Temperature-dependent gain and threshold in p-doped quantum dot
  lasers.
\newblock \emph{IEEE Journal of Selected Topics in Quantum Electronics},
  13:\penalty0 1261--1266, 2007.

\bibitem[Schowalter()]{Schowalter:priv_comm}
Marco Schowalter.
\newblock private communication.

\bibitem[nex()]{nextnano3}
\emph{nextnano$^{3}$ code, TU Munich (WSI), see
  http://www.nextnano.de/nextnano3/}.

\bibitem[Coldren and Corzine(1995)]{Coldren:95}
L.A. Coldren and S.W. Corzine.
\newblock \emph{Diode Lasers and Photonic Integrated Circuits}.
\newblock Wiley, New York, 1995.

\bibitem[Abdollahinia et~al.(2018)Abdollahinia, Banyoudeh, Rippien, Schnabel,
  Eyal, Cestier, Kalifa, Mentovich, Eisenstein, and
  Reithmaier]{Abdollahinia:18}
A.~Abdollahinia, S.~Banyoudeh, A.~Rippien, F.~Schnabel, O.~Eyal, I.~Cestier,
  I.~Kalifa, E.~Mentovich, G.~Eisenstein, and J.P. Reithmaier.
\newblock Temperature stability of static and dynamic properties of 1.55
  \text{mu}m quantum dot lasers.
\newblock \emph{Opt. Express}, 26\penalty0 (5):\penalty0 6056--6066, Mar 2018.
\newblock \doi{10.1364/OE.26.006056}.
\newblock URL \url{http://www.opticsexpress.org/abstract.cfm?URI=oe-26-5-6056}.

\bibitem[Lorke et~al.(2011)Lorke, Nielsen, and M{\o}rk]{Lorke:11}
M.~Lorke, T.~R. Nielsen, and J.~M{\o}rk.
\newblock Switch-on dynamics of nanocavity laser devices.
\newblock \emph{Applied Physics Letters}, 99\penalty0 (15):\penalty0 151110,
  2011.
\newblock ISSN 00036951.
\newblock \doi{DOI:10.1063/1.3651765}.
\newblock URL \url{http://dx.doi.org/10.1063/1.3651765}.

\end{thebibliography}

\end{document}